\documentclass[11pt, a4paper]{article} 

\usepackage[top=24mm, bottom=24mm, left=27mm, right=27mm]{geometry}

\usepackage{amsfonts}
\usepackage{amsthm, 
mathrsfs, 
mathtools, 
amsmath,
physics,
amssymb,
xcolor
}
\usepackage{bbm}
\usepackage{slashed}
\usepackage{mathalfa}
\usepackage{graphicx,caption,subfigure}
\usepackage{booktabs}
\usepackage{adjustbox}
\usepackage{scalefnt}
\usepackage{pifont}
\usepackage[utf8]{inputenc}
 \usepackage[splitrule]{footmisc}

\usepackage{placeins}
\usepackage{empheq}
\usepackage{paralist}
\usepackage{cite}
\usepackage[normalem]{ulem}
\usepackage{soul}

\usepackage[colorlinks=false,urlbordercolor=red]{hyperref}
\usepackage{xcolor}
\usepackage{tensor}
\usepackage{comment}

\usepackage{tikz,pgfplots}

\allowdisplaybreaks[2]
\numberwithin{equation}{section}

\begin{document}

\vspace*{-1.5cm}
\begin{flushright}
  {\small
  LMU-ASC 02/24\\ MPP-2024-6
  }
\end{flushright}

\vspace{1.5cm}
\begin{center}
{\Large \bf Minimal Black Holes and Species Thermodynamics} 

\end{center}

\vspace{0.35cm}
\begin{center}
{\large 
Ivano Basile$^a$,  Niccol\`o Cribiori$^b$, Dieter L\"ust$^{a,b}$ and Carmine Montella$^{b}$
}
\end{center}

\vspace{0.1cm}
\begin{center} 
\emph{
 $^a${\it Arnold Sommerfeld Center for Theoretical Physics,\\
Ludwig-Maximilians-Universit\"at M\"unchen, 80333 M\"unchen, Germany},   \\[0.1cm]
 \vspace{0.3cm} 
 $^b$Max-Planck-Institut f\"ur Physik (Werner-Heisenberg-Institut), \\[.1cm] 
   Boltzmannstrasse 8, 85748 Garching, Germany, 
   \\[0.1cm] 
    } 
\end{center} 

\vspace{0.5cm}

\begin{abstract}

The species scale provides a lower bound on the shortest possible length that can be probed in gravitational effective theories. It may be defined by the size of the minimal black hole in the theory and, as such, it has recently been given an interpretation along the lines of the celebrated black hole thermodynamics. In this work, we extend this interpretation to the case of charged species. We provide working definitions of minimal black holes for the case of uncharged and charged species constituents. Then, examining the modifications in the thermodynamic properties of near-extremal charged species compared to the uncharged case, we uncover interesting implications for the cosmology of an expanding universe, particularly within the context of the Dark Dimensions Scenario. Finally, we explore possible microscopic constructions in non-supersymmetric string theories in which towers of charged near-extremal species may arise.

\end{abstract}

\thispagestyle{empty}
\clearpage

\setcounter{tocdepth}{2}

\tableofcontents


\section{Introduction}

There are strong indications that gravity is holographic. As originally proposed by \cite{tHooft:1993dmi,Susskind:1994vu}, the holographic principle states that the description of a gravitational system inside a $d$-dimensional region of spacetime is encoded in its $(d-1)$-dimensional boundary.
This remarkable property is manifest in the celebrated result of Bekenstein and Hawking \cite{Bekenstein:1973ur,Hawking:1975vcx,Hawking:1976de} that the entropy of black holes is determined by the area law,
\begin{eqnarray}\label{Bekenstein}
{\cal S}_{BH}\simeq (R_{BH}M_P)^{d-2}\, .
\end{eqnarray}
Here, $R_{BH}$ denotes the radius of the event horizon of a Schwarzschild black hole and $M_P$ is the Planck mass in a $d$-dimensional gravitational theory.
The holographic nature manifest in this area law implies that the entropy determining the number of black hole degrees of freedom is not proportional to the volume of the black hole, as it would be expected for a large number of particles in a box, but rather to the area of the surface that surrounds it.
Another famous manifestation of holography in quantum gravity is the AdS/CFT correspondence, which states that quantum gravity in a $d$-dimensional anti-de Sitter space can be isomorphically mapped to a conformal field theory formulated on its $(d-1)$-dimensional conformal boundary \cite{Maldacena:1997re, Witten:1998qj}.

As discovered by Hawking, black holes are not absolutely stable, but when coupled to quantum field theory they can decay into particles \cite{Hawking:1975vcx}. 
In the semiclassical limit, the black hole decay spectrum is thermal and characterized by the Hawking temperature, $T_{BH}$, which is given by the surface gravity and thus 
is inversely proportional to the size of the event horizon,
\begin{eqnarray}
{T}_{BH}\simeq R_{BH}^{-1}\, .
\end{eqnarray}
This fact makes it clear that black holes are behaving similarly to thermodynamic systems and in fact obey laws of black hole thermodynamics. Specifically, the first law relates the change of black hole mass to that of the area or entropy, 
\begin{equation}\label{1stlaw2}
dM_{BH}=T_{BH}d {\cal S}_{BH}\, .
\end{equation}
The second law states that the horizon area, namely the black hole entropy, classically does not decrease as a function of time,
\begin{equation}\label{1stlaw3}
d{\cal S}_{BH}\geq0\, .
\end{equation}
Black hole thermodynamics also suggests the remarkable fact that the entropy could be understood from counting microstates forming a black hole of a given mass. 
This expectation was first confirmed by \cite{Strominger:1996sh}, for certain supersymmetric black holes constructed as configurations of $D$-branes in string theory.

Instead, the entropy of a system of particles in quantum mechanics is not expected to be given by an area law, but rather it is extensive, i.e.~proportional to the volume of the box in which particles are moving. 
The extensive entropy behavior is true for massless particles that are light compared to the ultraviolet (UV) cut-off scale, $\Lambda_{UV}$, of the low energy effective field theory (EFT) of quantum gravity.
However this behavior may change, especially if there are towers with a large number of particles and with masses up to $\Lambda_{UV}$ that can collapse and form a black hole.
Such towers of particle species occur naturally in quantum gravity, and this expectation is encoded in the swampland distance conjecture \cite{Ooguri:2006in} which predicts an infinite tower of light particles at infinite distance in moduli space. Additionally, according to the emergent string conjecture \cite{Lee:2019wij}, at the boundary of the moduli space, there are only two possibilities for a leading tower: light string excitations, signaling the emergence of a tensionless string, or Kaluza-Klein (KK) states, corresponding to decompactification to a higher-dimensional EFT.

In \cite{Cribiori:2023ffn}, it has been argued that the collective behavior of a tower of species is holographic and can be described in terms of {\it species thermodynamics}. This means that species carry entropy, temperature and also an overall total mass which precisely follow the rules of (black hole) thermodynamics. 
Indeed, the very reason why species thermodynamics can be formulated is arguably that species themselves are closely related to black holes. Indeed, in \cite{Cribiori:2023ffn, Basile:2023blg} this connection has been outlined comparing minimal black holes to ``typical'' ensembles of species, whose thermodynamic quantities match.\footnote{With ``typical'' we mean that momenta in some reference frame (corresponding to the rest frame of the associated minimal black hole) are of the order of the mass gap, while occupation numbers are order one.} 
This interconnection arises due to the species scale $\Lambda_{sp}$ \cite{Dvali:2007hz,Dvali:2007wp,Dvali:2009ks,Dvali:2010vm,Dvali:2012uq, Calmet:2014gya},  which provides an upper bound for $\Lambda_{UV}$ in any gravitational EFT.
In fact, $\Lambda_{sp}$ can be defined as (an upper bound on) the scale at which gravitational interactions become strong,
\begin{equation}\label{oldSC}
\Lambda_{sp} = \frac{M_{P}}{N_{sp}^{\frac{1}{d-2}}}\, ,
\end{equation}
where $N_{sp}$ is the number of light particles species. 
Note that this definition is implicit since $N_{sp}$ also depends on $\Lambda_{sp}$. An alternative, but related definition sets $\Lambda_{sp}$ as the scale at which higher-derivative operators in the gravitational EFT become relevant. 
The species scale and its moduli dependence have been recently subject of intense investigations
\cite{Cribiori:2021gbf,Castellano:2021mmx,Lust:2022lfc,Blumenhagen:2022dbo,Cota:2022yjw,Castellano:2022bvr,vandeHeisteeg:2022btw,Cota:2022maf,Cribiori:2022nke,Castellano:2023qhp,vandeHeisteeg:2023ubh,Andriot:2023isc,vandeHeisteeg:2023uxj,Cribiori:2023ffn,Cribiori:2023sch,Calderon-Infante:2023ler,vandeHeisteeg:2023dlw,Castellano:2023aum, Calderon-Infante:2023uhz,Lust:2023zql}.

The inverse of the species scale, the species length $L_{sp}=\Lambda_{sp}^{-1}$, sets the shortest possible length that can be resolved in any EFT before gravity becomes strongly coupled. 
When inverting the relation \eqref{oldSC} to get $N_{sp}\simeq (L_{sp}M_P)^{d-2}$, we see that $N_{sp}$ is following an area law, just as depicted in \eqref{Bekenstein} for black holes.
Based on this analogy, it was proposed in \cite{Cribiori:2023ffn} that $N_{sp}$ is indeed a non-extensive quantity playing the role of the species entropy ${\cal S}_{sp}$. 
More in general, the entropy of the tower of particle species can be defined as
\begin{equation}
{\cal S}_{sp} \simeq (L_{sp}M_P)^{d-2}\, ,
\end{equation}
up to additive log corrections. This proposal is in fact well-motivated by the observation that $L_{sp}$ corresponds to the radius of the smallest possible black hole that can be constructed as classical solution of the gravitational EFT.
This is deemed a {\sl minimal black hole}. 
Hence, the species entropy ${\cal S}_{sp}$ coincides with that of the minimal black hole, while the species temperature (for Schwarzschild-like black holes) is given by the inverse radius of the minimal black hole, namely
\begin{equation}\label{templambda}
T_{sp}=\Lambda_{sp}\, .
\end{equation}
It sets the highest possible temperature that can be obtained in the EFT.

In this paper, we extend the correspondence between species on the one side and minimal black holes on the other side, as well as the thermodynamic properties they share, studying (near-extremal) charged black holes. 
For this purpose, we start reviewing the N-portrait description \cite{Dvali:2011aa}, which we conveniently adapt to describe massive species particles as constituents. 
This description, although used only as a representation, will turn out to be beneficial for our purposes, since it will allow to compute entropy, temperature and also energy of species in an efficient way, and to compare these results with the appropriate thermodynamic laws, such as \eqref{1stlaw2}.

Another purpose of this paper is to compute the species temperature for various kinds of species. For electrically neutral species, corresponding to Schwarzschild-like black holes, the temperature in terms of the species scale is given by \eqref{templambda}. However, for charged and also spinning species this relation will be refined. Charged species correspond to charged minimal black holes, such as Reissner--Nordstr\"om solutions.
To derive their energy and temperature, we will extend the N-portrait picture to charged constituents, for both the extremal (BPS) and near-extremal (non-BPS) case (see also \cite{Gultekin:2023dif})
As a result, we will see that the temperature for near-extremal charged species will be reduced compared to Schwarzschild-like species and, instead of \eqref{templambda}, its parametric dependence in terms of $\Lambda_{sp}$ will be given by

\begin{equation}\label{templambda1}
T_{sp}=\Lambda_{sp}^{2}\, .
\end{equation}

The implications of this outcome find application in the scenario pertaining to the Dark Dimension \cite{Montero:2022prj}, where we show that temperature and decay rates of near-extremal species can nicely accommodate the properties of KK dark matter particles, as proposed in \cite{Gonzalo:2022jac, Obied:2023clp}. This result may underline the already recognized link between Dark matter graviton gases and black holes \cite{Anchordoqui:2022tgp}.

The computation of the species temperature is also useful to investigate the decay properties of towers of species and to predict their decay rate on rather general grounds.
We will discuss the thermal Hawking-like decay of species, whose rate is of the order of the species temperature. In addition, when the tower of species has departed from the semiclassical regime, we estimate the quantum decay rate which is dictated by the mass of the species particles and further suppressed compared to the thermal one.
This discussion can be also relevant in cosmology, to gain general information about species production and decay in an expanding hot universe.

\section{Minimal black holes as species bound states}

In this section, we study Schwarzschild-like black holes as bound states of species. 
Being governed by a single parameter, they are arguably the simplest class to be considered. 
Within the N-portrait picture, we propose a definition of minimal black holes and use it to define the species scale, entropy and temperature. 
Then, we comment on the duality between species as particles and as black hole geometries. 
Finally, we provide a microscopic counting of the entropy of these minimal Schwarzschild-like black holes supporting the thermodynamic interpretation of species.  

\subsection{The black hole N-portrait with massive species}

The black hole N-portrait  \cite{Dvali:2011aa,Dvali:2012rt,Dvali:2012uq,Dvali:2013eja} provides a quantum-mechanical description of black holes which does not rely on classical geometry. In this picture, black holes are described as bound states (Bose-Einstein condensates) of $N_g$ weakly-interacting massless gravitons with critical wave length $R_c = \sqrt{N_g}L_p$. Interactions among gravitons are controlled by the coupling
\begin{equation}
\label{alphag}
\alpha_g= \frac{L_P^2}{R_c^2}\simeq \frac{1}{N_g}\, ,
\end{equation}
which is weak when $N_g\gg 1$, namely when $R_c \gg L_P$. Since the number of ways in which the constituents can form a black hole is of order $n\simeq \exp N_g$, the entropy is
\begin{equation}
{\cal S}_{BH}\simeq N_g\, .
\end{equation}
The black hole radius and mass are given respectively by $R_{BH} = R_c \simeq \sqrt{N_g}L_P$ and $M_{BH} \simeq \sqrt{N_g}M_P$. 
Hence, the model is completely characterized by the parameter $N_g$. One can notice that the above relations give the species scale in a theory with $N_g$ massless species.

In addition to massless gravitons, we now include also a number $N_{sp}$ of massive particles species as part of the black hole constituents. We closely follow the discussion of \cite{Dvali:2012uq}.
Depending on whether these species couple non-gravitationally, such as quarks and leptons in the Standard Model, or if they couple gravitationally, such as towers of massive spin-2 KK gravitons, their effect on the black hole bound state is different. 
For the time being, we assume that species do not carry charges except for their masses. We thus concentrate on the case of gravitational species which are sourced by the energy momentum tensor, just like the graviton. 
Another justification for this assumption is that typical towers in string theory, namely KK or string towers \cite{Lee:2019wij}, always contain spin-2 particles and thus couple gravitationally.

In the presence of additional $N_{sp}$ species as constituents, a  black hole is formed when a modified quantum criticality condition is met, namely
\begin{equation}\label{critical}
\alpha_g= \frac{L_P^2}{R_c^2}\simeq \frac{1}{N_gN_{sp}}\, .
\end{equation}
In other words, the consequence of having additional species is that the effective total number of black hole constituents is now given by
\begin{equation}
N_{tot}=N_gN_{sp}\, ,
\end{equation}
meaning that the gravitational species contribute to $N_{tot}$ not additively but multiplicatively. This
is analogous to considering e.g.~two independent KK towers with $N_{sp,1}$ and $N_{sp,2}$ for which the total number of species is given by the product $N_{tot}=N_{sp,1}N_{sp,2}$ \cite{Castellano:2021mmx}.

We can determine how the black hole radius and mass are affected by the insertion of $N_{sp}$. 
From the relation \eqref{critical} and introducing the species length $L_{sp}=\sqrt N_{sp}L_P=1/\Lambda_{sp}$, it follows that the black hole radius $R_{BH}$ is
\begin{equation}
\label{BHRadius1}
R_{BH}=R_c\simeq \sqrt {N_gN_{sp}}L_P=\sqrt{N_g}L_{sp},
\end{equation}
while the mass is given by
\begin{equation}
M_{BH}\simeq \sqrt {N_gN_{sp}}M_P=\sqrt {N_g}L_{sp}/L_P^2\, .
\end{equation}
The number $n$ of ways in which the $N_{tot}$ constituents can form a black hole bound state is of order $n\simeq \exp N_{tot}$. Hence, at leading order in an expansion at large $N_{tot}$, the black hole entropy can be estimated as 
\begin{equation}
{\cal S}_{BH}\simeq N_gN_{sp}\, .
\end{equation}

Let us finally consider the loss of mass of black holes due to the thermal Hawking radiation. In the N-portrait picture, black holes decay via leakage of gravitons, which can occur due to the fact that the escape energy slightly exceeds the energy of the whole condensate. For large entropy, the corresponding decay rate $\Gamma_{BH}$, which in natural units is basically the associated Hawking temperature $T_{BH}$, can be computed in the
semiclassical approximation to be
\begin{equation}
\Gamma_{BH}\simeq T_{BH}\simeq \frac{1}{\sqrt{N_gN_{sp}}L_P}=\frac{1}{\sqrt {N_g}L_{sp}}\, .
\end{equation}

The N-portrait picture including massive gravitational species can be easily generalized to Schwarzschild-like black holes in arbitrary $d$ spacetime dimensions.  
Concretely, one can derive the following relations between mass, temperature, Schwarzschild radius and entropy (in $d$-dimensional Planck units):
\begin{eqnarray}
\label{bhrel}
M_{BH}&=&\bigl(R_{BH}\bigr)^{d-3}={\cal S}_{BH}^{\frac{d-3}{d-2}}\, ,\nonumber\\
T_{BH}&=&\bigl(R_{BH}\bigr)^{-1}={\cal S}_{BH}^{-\frac{1}{d-2}}\, .
\end{eqnarray}
This is in agreement with the first law of black hole thermodynamics,
\begin{equation}\label{1stlaw}
dM_{BH}=T_{BH}\, d {\cal S}_{BH}\, ,
\end{equation}
leading to
\begin{equation}
\frac{1}{T_{BH}}=\frac{\partial {\cal S}_{BH}}{ \partial M_{BH}}\, .
\end{equation}

Let us emphasize that the above thermodynamic relation between black hole temperature and entropy is only true in the semiclassical approximation, i.e.~in the asymptotic regime ${\cal S}_{BH}\rightarrow\infty$.
For small black holes, and in particular after the decay of roughly half of the black hole, additional quantum effects become relevant. They will result in a non-thermal black hole decay whose rate will be slower than that of the thermal Hawking radiation.
If we parametrize the quantum suppression of the black hole decay with a power law dependence on the entropy, we are led to a quantum decay rate \cite{Dvali:2020wft}
\begin{equation}
\Gamma_{BH,qm}\simeq \frac{T_{BH}}{ {\cal S}_{BH}^n}\simeq {\cal S}_{BH}^{\frac{1-n(2-d)}{ 2-d}}\,.
\end{equation}
At the moment, we simply assume for Schwarzschild-like black holes that $n\geq 0$. Later, we will try to determine $n$ for different particle species.

\subsection{Definition of minimal black hole, species entropy and temperature}

We now describe how to construct minimal black holes as bound states of species. 
The size of the minimal black hole corresponds to the minimal pixel that can be resolved in any effective theory of gravity, namely to the shortest possible length  \cite{Dvali:2012uq},
\begin{equation}
R_{BH,min}\simeq L_{sp}\, .
\end{equation}
Hence, the minimal black hole in the N-portrait picture is the bound state of gravitons and species in which the number of massless graviton constituents is set to its minimal value, i.e.~one: $N_g = 1$.

Effectively, the number $N_g$ of massless gravitons of the original N-portrait picture without species is replaced by $N_{sp}$, the number of massive species modes.
In general, the minimal black hole constituents do not have all equal mass, but may form a tower of states with a certain mass spacing.

Following this prescription for the entropy, mass and temperature of the minimal black hole in $d$-dimensions we obtain respectively:
\begin{eqnarray}
{\cal S}_{BH,min}&\simeq& N_{sp}\, ,\\
M_{BH,min}&\simeq &N_{sp}^{\frac{d-3}{d-2}}M_P\, ,\\
T_{BH,min}&\simeq &N_{sp}^{{-\frac{1}{d-2}}}M_P\, .
\end{eqnarray}
Any other black hole in the effective field theory will have radius, mass and entropy equal or larger than those above,
\begin{equation}
{\cal S}_{BH}\geq {\cal S}_{BH,min},\qquad R_{BH}\geq R_{BH,min},\qquad M_{BH}\geq M_{BH,min}\, .
\end{equation}

Having provided a working definition of minimal black hole, we can now employ it to define the various quantities associated to the thermodynamics of species. Given a tower of $N_{sp}$ particle species, we define its entropy and mass to be that of the associated minimal black hole, namely
\begin{align}
{\cal S}_{sp} \equiv {\cal S}_{BH,min},\\
M_{sp} \equiv M_{BH,min}\, . 
\end{align}
In turn, this leads to the following expressions
in terms of the species scale $\Lambda_{sp}$:
\begin{eqnarray}
{\cal S}_{sp}&\simeq& \Lambda_{sp}^{2-d}\, ,\\
M_{sp}&\simeq& \Lambda_{sp}^{3-d}\, .
\end{eqnarray}

Similarly to black holes, species particles are in general not stable, but they can interact among themselves and with other particles. These interactions will lead to a decay of the non-stable (non-BPS) species, in analogy to the decay of the corresponding minimal black hole. 
Due to this decay, the species tower will loose mass and entropy.
For species corresponding to Schwarzschild-like black holes, the decay rate in the limit of large entropy is dominated by the semiclassical Hawking decay rate of the associated minimal black hole. 
In this case, independently from the microscopic nature of the interactions among species, one can derive a temperature for the particle species,
\begin{equation}\label{tsp}
T_{sp}\equiv T_{BH,min}\simeq \Lambda_{sp}\, .
\end{equation}
Hence, the temperature of any black hole in the effective theory must be equal or smaller than the species temperature,
\begin{equation}
T_{BH}\leq T_{sp}\, .
\end{equation}

Note that the species entropy and temperature obey the same relations as their black hole companions, namely in $d$-dimensions and for the case Schwarzschild-like case we have that
\begin{eqnarray}\label{tsp1}
T_{sp}={\cal S}_{sp}^{-\frac{1}{d-2}}\, .
\end{eqnarray}
We emphasize again that this temperature is relevant for the decay of the species in the semiclassical approximation with ${\cal S}_{sp}\rightarrow\infty$. Once roughly half of the species have decayed, the
semiclassical approximation breaks down and quantum effects become relevant. As a result, the decay rate of the species is reduced.
Following the analogous discussion on black holes, we assume that the quantum decay rate of species can be parametrized as
\begin{equation}\label{eq:qm_decay_rate}
\Gamma_{sp,qm}\simeq \frac{T_{sp}}{{\cal S}_{sp}^n}\simeq {\cal S}_{sp}^{\frac{1-n(2-d)}{2-d}}\simeq \Lambda_{sp}^{1+n(d-2)}\,.
\end{equation}
Let us now try to determine the value of the suppression parameter $n$ of the species decay. 
We assume that for large ${\cal S}_{sp}$ the species originally possess a thermal distribution as given by the temperature $T_{sp}$ in \eqref{tsp} and \eqref{tsp1}.
Therefore, this distribution is peaked at particles species with mass
\begin{equation}
\langle m_{sp} \rangle\simeq T_{sp}\simeq \Lambda_{sp}\, .
\end{equation}
We can justify this as follows. In $d$ external spacetime dimensions plus $q$ extra dimensions, consider a KK tower of mass scale $m$ at equilibrium with temperature $T \gg m$. 
Neglecting interactions, and up to a volume prefactor which will cancel when considering the Boltzmann distribution, the partition function can be approximated by the classical expression 
\begin{equation}
Z_\text{KK} = \sum_{\vec{k} \in \mathbb{Z}^{q}-\{0\}} \int \frac{d^{d-1}p}{(2\pi)^{d-1}} \, e^{- \frac{1}{T}\sqrt{p^2+m^2k^2}} \propto \sum_{\vec{k} \in \mathbb{Z}^{q}-\{0\}} \left(\frac{m}{T} \abs{k}\right)^{\frac{d}{2}} K_{\frac{d}{2}}\left(\frac{m}{T} \abs{k}\right).
\end{equation}
Here, $\abs{k} \equiv \sqrt{k \cdot k}$ is the norm of the lattice vector $\vec{k} \in \mathbb{Z}^q$, while $K_\nu(z)$ the modified Bessel function. 
The probability distribution for $k$ is just 
\begin{equation}
p(k) = \frac{\left(\frac{m}{T} \abs{k}\right)^{\frac{d}{2}} K_{\frac{d}{2}}\left(\frac{m}{T} \abs{k}\right)}{\sum_{\vec{l} \in \mathbb{Z}^{q}-\{0\}} \left(\frac{m}{T} \abs{l}\right)^{\frac{d}{2}} K_{\frac{d}{2}}\left(\frac{m}{T} \abs{l}\right)} \, ,
\end{equation}
and thus the distribution of the species masses, $m_k = m \abs{k}$, for large $|k|$ is approximately proportional to 
\begin{equation}
\abs{k}^{q-1} \left(\frac{m}{T} \abs{k}\right)^{\frac{d}{2}} K_{\frac{d}{2}}\left(\frac{m}{T} \abs{k}\right) \overset{\abs{k} \gg 1}{\sim} m_k^{q-1+\frac{d-1}{2}} \, e^{- \frac{m_k}{T}} \, ,
\end{equation}
where we also provided the large-mass (equivalently low-temperature) asymptotics using that $K_\nu(z) \sim \sqrt{\frac{\pi}{2z}}e^{-z}$ for $z\gg 1$. The latter expression shows the expected Boltzmann suppression factor.
For $q>1$ extra dimensions there is indeed a peak for an order-one value of the argument, namely at $m_k \sim T$. For $q=1$ extra dimensions there is no peak in the function $z^\frac{d}{2}K_{\frac d2}(z)$, but the expectation value of $m_k$ can be still estimated: for $m \ll T$, the (unnormalized) expectation value of a generic power $k^\gamma$ is given by
\begin{equation}
\sum_{k>0}  k^\gamma K_{\frac{d}{2}}\left(\frac{m}{T}k\right) \sim \left(\frac{T}{m}\right)^\gamma \int_0^{N_{sp} m/T = \Lambda_{sp}/T} du \, u^\gamma \, K_{\frac{d}{2}}(u) \, ,
\end{equation}
which for $T \lesssim \Lambda_{sp}$ is of order $(T/m)^\gamma$, since the integral converges even when the upper integration limit is taken to infinity. The properly normalized expectation value of $\abs{k}$ for $q=1$ is thus a ratio of these powers, with $\gamma = 1$ and $\gamma = 0$, hence
\begin{equation}
\langle m_k \rangle = m \langle \abs{k} \rangle \sim m \, \frac{T}{m} = T \, .
\end{equation}
Summarising, for $T \sim \Lambda_{sp}$ one finds the expected behavior both with and without a peak in the distribution.

After half of the species have decayed, i.e.~for finite ${\cal S}_{sp}$, the species start to behave like  particles. We can then use a standard single particle description and, following simple dimensional arguments (otherwise, see e.g.~\cite{Giudice:1998ck,Han:1998sg}), the quantum-mechanical decay rate of a particle with mass $m_{sp}$ is approximately given in Planck units as
\begin{equation}
\Gamma_{sp,qm}\simeq m_{sp}^{d-1}\simeq T_{sp}^{d-1}\simeq \Lambda_{sp}^{d-1}\, .
\end{equation}
In terms of entropy suppression this corresponds to $n=1$ in eq.~\eqref{eq:qm_decay_rate}, which is compatible with the analysis of \cite{Dvali:2020wft}.

\subsection{Duality between species particles and black hole geometry}

The above definitions of species entropy and temperature as those of the associated minimal black hole can be interpreted in terms of a duality between particles and geometry: the single species being particles, while their collective behaviour corresponds to a particular space-time geometry.
More precisely, the entropy of a large number of particle species is identical to, and in fact inherited from, that of the corresponding minimal black hole. 
In this regime, the species entropy is not an extensive quantity with a volume scaling behavior, but rather it follows an area law, just like that of the associated minimal black hole. 
Furthermore, the decay rate of a large number of particle species is collectively determined by the thermodynamic species temperature.
On the other hand, a small numbers of species behave like quantum-mechanical particles and decay according to the rules of quantum mechanics. 
Hence, when the number of species decreases beyond a certain critical value, there will be a transition from a thermodynamic to a particle-like description. 
Additionally, this picture is linked to the idea of spacetime as a condensate or bound state of some generic quantum gravity microstates.

This kind of particle/geometry duality is consistent with the definition of species scale as the UV cut-off scale at which gravity becomes strongly coupled. 
In fact, it is natural to assume that at strong coupling species are forming the minimal black holes as 
bound state, such that there is a transition between the particle and the geometric description. 
Summing over the gravitational species in a loop diagram provides the higher curvature corrections to the Einstein action, whose strengths are controlled by the species scale \cite{vandeHeisteeg:2022btw,Cribiori:2022nke,vandeHeisteeg:2023ubh,Cribiori:2023sch}. This is also  related to the idea of emergence, namely that gravity and strings can emerge when integrating out towers of species particles \cite{Heidenreich:2017sim,Grimm:2018ohb,Palti:2019pca,Castellano:2022bvr,Blumenhagen:2023yws,Blumenhagen:2023tev,Blumenhagen:2023xmk,Calderon-Infante:2023uhz}.

The above picture is also closely related to the Black Hole Entropy Distance Conjecture \cite{Bonnefoy:2019nzv}, which states that in the limit of large black hole entropy, ${\cal S}_{BH}\rightarrow\infty$, there is a tower of light states, which can be related to black hole microstates \cite{Bonnefoy:2019nzv} with energies $m$ such that
\begin{equation}
{\cal S}_{BH}\rightarrow\infty\qquad\Rightarrow \qquad m\sim\left(\mathcal{S}_{BH}\right)^{-\gamma}\rightarrow 0\, ,\quad \gamma>0\, .
\end{equation}
For the entropy of the minimal black hole, i.e.~for the species entropy, there is an analogous behaviour, namely
\begin{equation}
{\cal S}_{sp}\rightarrow\infty\quad\Rightarrow\quad m_{tower}\sim\left(\mathcal{S}_{sp}\right)^{-\gamma}\rightarrow 0\, .
\end{equation}
Now, the light states are the species themselves and $m_{tower}$ is the mass scale of the tower which may be seen as lightest energy fluctuation of a minimal black hole (namely $m_{KK}$ or $M_s$ for KK or string towers).

\subsection{Kaluza-Klein species bound states as minimal black holes}

The thermodynamic picture of species can be supported with a direct counting argument, in both the particle and the black hole representation of species. We start from the former and then look at the latter.

For concreteness, we consider a tower of species such as
\begin{equation}
M_{n} = n^\frac{1}{p}\,\Delta m,
\end{equation}
where $\Delta m$ is a mass spacing that we need not to specify further and $p$ is a parameter modelling the density of the tower. For example, one has $p=1$ for a single KK tower, while one takes $p\to\infty$ to extrapolate the result for a string tower \cite{Castellano:2021mmx}. The number of species is then 
\begin{equation}
\label{Nsp_ptower}
N_{sp} = \left(\frac{\Lambda_{sp}}{\Delta m}\right)^p.
\end{equation}
As computed in \cite{Cribiori:2023ffn}, the total mass of the tower up to $n=N_{sp}$ is given by the generalized harmonic number $H_n^{(q)}$,
\begin{equation}
M_{sp} =\sum_{n=1}^{N_{sp}}M_{n} =\Delta m \, H_{N_{sp}}^{(-\frac 1p)}.
\end{equation}
By looking at the asymptotic behaviour for $N_{sp}\gg 1$, we find
\begin{equation}
M_{sp} \simeq \frac{p}{p+1}\Lambda_{sp} N_{sp} \simeq \frac{p}{p+1} M_p N_{sp}^{\frac{d-3}{d-2}}.
\end{equation}
Hence, the dependence of the total mass of the tower on $N_{sp}$ coincides with that of the mass of the black hole, up to unimportant order one factors which we are not keeping track of. This conclusion holds also for $p \to \infty$, namely for a string tower, supporting the thermodynamic interpretation within the picture of species as particles \cite{Basile:2023blg, Cribiori:2023ffn}.

Next, we repeat the same calculation in the black hole picture. 
The idea is to count the number of microstates in a black hole, which we assume to be Schwarzschild-like and formed out of $N_{sp}$ species as in \eqref{Nsp_ptower}. 
This has been done in \cite{Blumenhagen:2023yws} for $p=1$ and here we generalize it to generic $p$.
Given the black hole mass $M_{BH}$ and the scale $\Delta m$, we can define the number
\begin{equation}
N = \frac{M_{BH}}{\Delta m},
\end{equation}
which counts the total mass level associated to the black hole. This level can be reached combining (multi)particle states with levels up to $N_{sp}$. For a minimal Schwarz\-schild-like black hole, we have then
\begin{equation}
\label{NNsprel}
N  = \frac{M_{p}^{d-2}}{\Lambda_{sp}^{d-2}} \frac{\Lambda_{sp}}{\Delta m} = N_{sp}^{\frac{p+1}{p}}.
\end{equation}
We want to find that the entropy of the black hole is given by $N_{sp}$ and not by $N\gg N_{sp}$ as one might have naively expected. 
By definition, the entropy is the logarithm of the number of microstates. 
These are in our case related to the number of partitions of $N$ into $N_{sp}$ parts.\footnote{More precisely, we should consider the number of partitions of $N$ with largest part $N_{sp}$. A classic result in combinatorics (see e.g.~\cite{flajolet_sedgewick_2009}, Example I.7) shows that this is equal to the number of partitions of $N$ into $N_{sp}$ parts.}. 
To find the appropriate relation, we start from the number of (weak) compositions of $N$ into $N_{sp}$ parts, $C_{N,N_{sp}}$, which is given by
\begin{equation}
C_{N,N_{sp}} = \left(
\begin{array}{c}
N + N_{sp}-1\\
N
\end{array}
\right) \sim  \frac{N^{N_{sp}-1}}{(N_{sp}-1)!},
\end{equation}
where in the last steps we wrote the asymptotic behaviour for $N\gg N_{sp}$. This regime is justified in our case due to \eqref{NNsprel}.
Compositions differing by the order of their elements are considered the same partition. Hence, to get the appropriate number of partitions we should divide by the number of compositions which are equal up to ordering. 
For $p=1$, this is $N_{sp}!$ and one recovers the expression used in \cite{Blumenhagen:2023yws}. 
However, for generic $p>1$ the factor $N_{sp}!$ overestimates the degeneracy. 
To understand this, notice that we can equivalently think of \eqref{Nsp_ptower} as the number of species coming from $p$ different KK-towers. When elements of two partitions are the same but arise from different towers, we consider the partitions to be different. In other words, a multiparticle state with mass $2\Delta m_1$ is considered different from another with mass $\Delta m_1+\Delta m_2$, even for $\Delta m_1 = \Delta m_2$. 
Due to this fact and assuming towers to be homogeneously populated, since the quantum numbers of multiple towers multiply we estimate that the degeneracy in the number compositions with $p$ different towers is reduced from $N_{sp}!$ down to $(N_{sp}!)^\frac{1}{p}$. 
Hence, we calculate the entropy as
\begin{equation}
e^{S_{BH}} \simeq \frac{C_{N,N_{sp}}}{(N_{sp}!)^\frac{1}{p}} \sim \frac{N^{N_{sp}-1}}{(N_{sp}!)^\frac{1}{p}(N_{sp}-1)!}\sim \frac{N^{N_{sp}}}{(N_{sp}!)^\frac{p+1}{p}},
\end{equation}
where we used that $N_{sp}\gg 1$. Employing the relation \eqref{NNsprel} between $N$ and $N_{sp}$, we find
\begin{equation}
{S_{BH}} \sim \log \frac{N_{sp}^{\frac{p+1}{p}N_{sp}}}{(N_{sp}!)^\frac{p+1}{p}} \sim  {\frac{p+1}{p}N_{sp}}
\end{equation}
and thus $S_{BH} \simeq N_{sp}$ as desired, even when extrapolating the result for $p\to \infty$. This combinatorial estimate agrees with the microcanonical result obtained in \cite{Basile:2023blg}.

To summarize, the above computation confirms the thermodynamic interpretation for species both in the case in which these are particles-like or in which they are constituents of Schwarzschild-like black holes. We found once more that the total species mass corresponds to the mass of the minimal black hole and is related to the species entropy as
\begin{equation}
M_{sp,KK}=M_{BH,min}=  {\cal S}_{sp,KK}^{\frac{d-3}{ d-2}} =\Lambda_{sp}^{3-d}\, ,
\end{equation}
while the species temperature is 
\begin{equation}\label{tspkk}
T_{sp,KK}=R_{BH,min}^{-1}=  {\cal S}_{sp,KK}^{\frac{1}{2-d}} =\Lambda_{sp}\, .
\end{equation}

\section{Minimal charged black holes}
\label{sec:minchargedBH}

In this section, we study minimal charged black holes and their species thermodynamics. In particular, we give a recipe to construct minimal charged black holes via the attractor mechanism and also as bound states of massive, charged species.
As we are going to explain, some of the thermodynamic relations previously studied in the uncharged case are going to be revisited. 
In particular, the parametric dependence between the Hawking temperature and entropy will change, with the result that the semiclassical decay rate of the charged species is in general slower than that of uncharged ones.

\subsection{Extremal and non-extremal charged black holes}

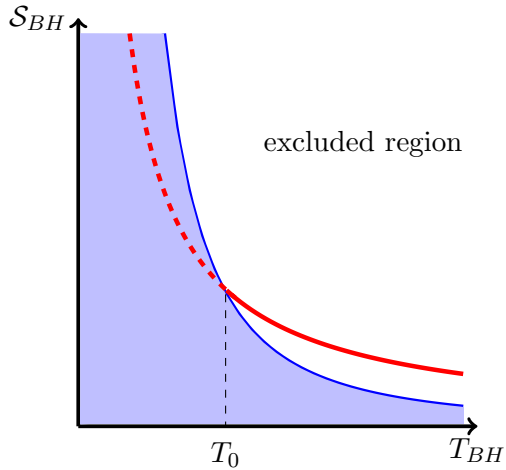
\begin{figure}[h]
\begin{center}
\begin{tikzpicture}[scale=1.25]
  \draw[->, line width=.5mm] (0, 0) -- (4.2, 0) node[below] {${T}_{BH}$};
  \draw[->,line width=.5mm] (0, 0) -- (0,4.33) node[left] {${\cal S}_{BH}$};
\draw[scale=1.5,domain=0.6:2.7,smooth, variable=\x,blue, thick,line width=.6mm] plot ({\x},{1/((\x)*(\x))});
\fill [blue!25, domain=0.1:4.2, variable=\x]
  (0.02, 4.17)
  -- plot[scale=1.5,domain=0.6:2.7] ({\x},{1/((\x)*(\x))})
  -- (0.02, 0.02);
  \fill [blue!25] (0.02,0.02) rectangle (4.05,0.2); 
  \node () at (3,3) {excluded region};
  \draw[scale=1.5,domain=1.04:2.7,smooth, variable=\x,red,line width=.6mm] plot ({\x},{1/(\x)});
  \draw[scale=1.5,domain=0.36:1.1,smooth, variable=\x,red, dashed,line width=.6mm] plot ({\x},{1/(\x)});
  \draw[dashed] (1.55,1.4)--(1.55,0);
  \node () at (1.55,-.3) {${ T}_0$};
\end{tikzpicture}
\end{center}
\caption{Parameter space for Reissner--Nordstr\"om black holes. 
Schwarzschild black holes (blue line, ${\cal S}_{BH}=1/T_{BH}^2$) are at the boundary between the allowed region ($Q_{BH}^2>0$) and the excluded one ($Q_{BH}^2<0$). The red line is the line of constant extremality parameter $c= \mathcal{S}_{BH} {T}_{BH}$. For ${T}_{BH}>{T}_0={1/c}$, one cannot keep $c$ constant while staying in the allowed region. The figure is taken from \cite{Cribiori:2022cho}.}
\label{fig:BHpopulation}
\end{figure}

We now deal with a multi-parameter family of black hole metrics, characterized by mass $M_{BH}$, together with electric and magnetic charges $Q_{BH,i}$ and $P_{BH}^i$. 
To exemplify our discussion, let us consider the Reissner--Nordstr\"om black hole in Einstein--Maxwell gravity, which is specified by mass $M_{BH}$ and electric charge $Q_{BH}$. Its entropy is given by the radius $R_{BH,+}$ of the outer horizon,
\begin{equation}
{\cal S}_{BH}\simeq \frac{R_{BH,+}^2}{L_P^2}\, .
\end{equation}
To define the temperature, it is convenient to introduce the extremality parameter\footnote{We will work in Planck units and be cavalier on unimportant numerical factors.}
\begin{equation}
c=\sqrt{M_{BH}^2-Q_{BH}^2}\, ,
\end{equation}
namely $R_{BH,\pm} =M_{BH}\pm c$.
Demanding it to be real, one gets the extremality bound for charged black holes\footnote{Later on we will discuss a refinement of the extremality bound due to the presence of higher-derivative corrections to the Einstein-Maxwell action.}
\begin{equation}
M_{BH}^2\geq Q_{BH}^2\, 
\end{equation}
and extremal black holes correspond to $c=0$.
In the presence of charges, the first law of black hole thermodynamics \eqref{1stlaw} is modified in the following way 
\begin{equation}
 dM_{BH} = T_{BH} d{\cal S}_{BH} + \Phi dQ_{BH},
\end{equation}
hence
\begin{equation}
d(M_{BH} - \Phi Q_{BH}) = T_{BH}d\mathcal{S}_{BH} - Q_{BH} d\Phi = dc,
\end{equation}
where $\Phi=Q_{BH}/R_{BH,+}$ is the electric potential at the black hole horizon. From this relation we can read off the temperature and the charge as
\begin{equation}
 T_{BH} = \frac{\partial c}{\partial {\cal S}_{BH}}\Big|_\Phi, \qquad Q_{BH} = -\frac{\partial c}{\partial \Phi}\Big|_{{\cal S}_{BH}}.
\end{equation}
Therefore, the black hole temperature can be expressed in terms of $c$ and the entropy as 
\begin{equation}
T_{BH}\simeq \frac{c}{\mathcal{S}_{BH}}\, .
\end{equation}
We see that for finite entropy extremal black holes have zero temperature. The reverse case is also possible, namely small black hole with vanishing entropy but finite temperature have $c=0$.
Compared to neutral black holes, charged black holes have always a smaller temperature and therefore their semiclassical decay rate is reduced as well, compared e.g.~to the Schwarzschild case.
The physical region of the parameter space for the Hawking temperature and entropy is shown in the in Figure \ref{fig:BHpopulation}, which is taken from \cite{Cribiori:2022cho}.

As discussed before, on top of the semiclassical reduction of the Hawking temperature, after half of their life time (charged) quantum black holes undergo yet another suppression of their decay rate. We then write the quantum decay rate of charged black holes as
\begin{equation}
\label{qmdecayRN}
\Gamma_{BH,qm}\simeq \frac{T_{BH}}{ {\cal S}_{BH}^n}\simeq ({\cal S}_{BH})^{-1-n}\,,
\end{equation}
for some parameter $n$ which we are going to determine in the following.

\subsection{Moduli-dependent species entropy and temperature}

Before introducing the definition of minimal charged black holes as bound states of charged species, we would like to give a recipe to extract the moduli dependence of the species entropy which is particularly well-suited for charged black holes in string theory. 
This method has already been employed in \cite{Cribiori:2022nke},  particularly in the presence of certain higher-derivative curvature corrections which source additional contributions to the entropy, 
besides the Bekenstein--Hawking part, in the form of a topological string free energy.

Consider the entropy at the horizon, $\mathcal{S}_{BH} = \mathcal{S}_{BH}(Q,P)$, as a function of the electric and magnetic charges $Q$, $P$. For definiteness, we have in mind a 1/2-BPS black hole in ${\cal N}=2$ supergravity, but the procedure can possibly be extended to other setups. In order to construct a minimal charged black hole, one can perform the following steps: 
\begin{itemize}

\item Set the electric or magnetic charges to minimal values,
\begin{equation}
P=P_{min}\qquad{\rm or}\qquad Q=Q_{min}\, .
\end{equation}
This is typically constrained by charge quantization and by the requirement of having a non-vanishing entropy. 

\item Replace the remaining charges in terms of the moduli $\phi$ via the attractor equations,
\begin{equation}
Q=Q(\phi) \qquad \text{or} \qquad P = P(\phi), \qquad \text{such that} \qquad \mathcal{S}_{BH} = \mathcal{S}_{BH}(\phi).
\end{equation}

\item Check that the Black Hole Entropy Distance Conjecture \cite{Bonnefoy:2019nzv} is satisfied at the boundary of the moduli space,
\begin{equation}
{\cal S}_{BH}(\phi)\rightarrow\infty \qquad {\rm for}\quad\phi\rightarrow\infty\, .
\end{equation}
If this is not the case, restart from the first point and change the choice of minimal charges; e.g.~if magnetic charges were minimized, try again by minimizing electric ones.

\end{itemize}

Let us briefly illustrate this procedure with a simple and well-known example of a non-extremal black hole solution in ${\cal N}=2$ supergravity.
The solution is realized in type IIA string theory via a Calabi-Yau compactification with a system of $Q$ $D0$-branes and three $D4$-branes wrapped $P_1$, $P_2$, $P_3$ times around Calabi-Yau 4-cycles.
To simplify the setup further, we set the magnetic charges equal to each other, $P_1=P_2=P_3\equiv P$.
The volume ${\cal V}_2$ of a 2-cycle is parametrize by a K\"ahler modulus and corresponds to the smallest possible homological cycle on a simply connected Calabi-Yau.  
When performing a large 2-cycle volume limit at fixed weak string coupling, the number of KK modes is proportional to $\mathcal{V}_2$, namely $N_{KK}\simeq \mathcal{V}_2$. The linear power can be fixed by using that the species scale in decompactification limits is the higher dimensional Planck scale.\footnote{Notice that we can also take a large volume, strong coupling limit. In particular, if we let $g_s\sim\mathcal{V}_2^{\frac32}$ in order to move just within the vector multiplet moduli space, then the eleventh dimensions of M-theory opens up, $g_s \sim (R_{11}M_{p,11d})^\frac32$. Then, we can identify precisely the KK species with a tower of D0-branes.}
 We want to recover this result by following the steps outlined above.

The lowest order black hole entropy, i.e.~without higher curvature contributions, and the extremality parameter $c$  are related in the following way \cite{Galli:2011fq, Cribiori:2022cho}\footnote{Up to numerical factors, the formula \eqref{relnonext} can be found from (4.44) of \cite{Cribiori:2022cho} after identifying the volume at the horizon with that at infinity, $\mathcal{V}_h \equiv \mathcal{V}_\infty$. This can be accomplished by tuning the values of the scalars at infinity and is in accordance with the Black Hole Entropy Distance Conjecture, since we are in a regime in which the black hole is growing in size. Notice that we fixed a typo in the analogous  formula (4.4) of \cite{Cribiori:2023ffn}.}
\begin{equation}\label{relnonext}
{\cal S}_{BH}\simeq(\sqrt{{\cal V}_2P^2+c^2}+c)^2\,.
\end{equation}
As a check, one can see that this expression reduces to the known one, ${\cal S}_{BH}\simeq\sqrt{QP^3}$, in the extremal case, where also $\mathcal{V}_2 \simeq \sqrt{Q/P}$. According to the above procedure, we set the magnetic charge to the minimal value $P=1$
and we get
 \begin{equation}
{\cal S}_{sp}\simeq(\sqrt{{\cal V}_2+c^2}+c)^2\,.
\end{equation}
Assuming ${\cal V}_2\gg c^2$, which is natural at large volume, the entropy of the minimal black hole becomes
\begin{equation}\label{CYVolumeBH}
{\cal S}_{sp}={\cal S}_{BH,min}  = \mathcal{V}_2 \simeq N_{KK} .
\end{equation}
Hence, we recovered that the species entropy is governed by the volume of a 2-cycle, which is a moduli-dependent proxy for number of KK particles.

We can finally derive the species temperature
 \begin{equation}
{T}_{sp}\simeq \frac{c}{{\cal S}_{sp}}\simeq {c\Lambda_{sp}^2}\,.
\end{equation}
In complete analogy with the previous discussion on charged (minimal) black hole, for fixed $c$ it scales like $T_{sp}\sim{\cal S}_{sp}^{-1}$ (namely along the dashed curve in fig. \ref{fig:BHpopulation}).
As for the quantum mechanical decay rate of the species tower after half of its life time, assuming again that the species distribution is peaked at a mass scale $m_{sp}\simeq T_{sp}$, we get that (for $d=4$)
\begin{equation}
\Gamma_{sp,qm}\simeq T_{sp}^{3}\simeq \Lambda_{sp}^{6}\, .
\end{equation}
When compared with \eqref{qmdecayRN}, this corresponds to the value $n=2$.

\subsection{Minimal charged black holes as species bound states}
\label{sec:chBHspecies}

In analogy with the discussion on Schwarzschild-like black holes, we would like to construct minimal charged black holes as bound states of charged species. 
Hence, let us consider a tower of states with masses and charges $(m_n,q_n)$, with $n=1,\dots , N_{sp}$. 
We assume that at each level the species particles respect the tower weak gravity conjecture \cite{Heidenreich:2015nta} (see \cite{Cota:2023uir} for a recent and careful analysis), which means that the tower is super-extremal, namely
\begin{equation}
|q_n|\geq m_n\quad \forall~n\, .
\end{equation}
The total mass and charge of the minimal black black, being the same as the total species mass and charge,  are then given as
\begin{equation}
Q_{BH,min}\equiv Q_{sp}=\sum_{n=1}^{N_{sp}}q_n \, ,\qquad M_{BH,min}\equiv M_{sp}=\sum_{n=1}^{N_{sp}}m_n\, .
\end{equation}
Assuming that all $q_n$ have the same sign, it follows that the minimal black hole is also super-extremal, i.e.~$|Q_{BH,min}|\geq M_{BH,min}$.
At first sight, this seems to violate the extremality bound for black holes. However, as discussed in \cite{Arkani-Hamed:2006emk, Kats:2006xp, Hamada:2018dde, Cheung:2018cwt, Charles:2019qqt} for small black holes, of which minimal black holes are an example, the extremality bound gets changed due to higher-derivative terms in the effective action. In our case, the relevant four-derivative operators are
\begin{equation}
\alpha_1(F_{\mu\nu}F^{\mu\nu})^2\, ,\qquad \alpha_3F_{\mu\nu}F_{\rho\sigma}W^{\mu\nu\rho\sigma}\, ,
\end{equation}
where $F^{\mu\nu}$ is the electromagnetic field-strength and $W^{\mu\nu\rho\sigma}$ is the Weyl tensor. The $\alpha_i$ are dimensionless Wilson coefficients in Planck units, in the notation of \cite{Hamada:2018dde}.
In the presence of these higher curvature corrections, the black hole extremality bound is changed as \cite{Kats:2006xp,Hamada:2018dde}
\begin{equation}
\frac{Q_{BH,min}}{ M_{BH,min}}\leq 1+ \frac{\alpha}{Q_{BH,min}^2}\, , \qquad \alpha={\frac25}(4\pi)^2(2\alpha_1-\alpha_3).
\end{equation}
Here, we are working in $d=4$ for concreteness.
Black holes saturating this bound are extremal, otherwise they are non-extremal.
It follows that also the extremality parameter gets modified in the presence of these corrections and it takes the form
\begin{equation}
\label{cextrhd1}
c^2=M_{BH,min}^2-Q_{BH,min}^2+\frac{2\alpha M_{BH,min}^2}{ Q_{BH,min}^2}+\frac{\alpha^2 M_{BH,min}^2}{ Q_{BH,min}^4}\, .
\end{equation}
For large charges, one can neglect the last term in the above equation. Assuming furthermore that $M_{BH,min}^2\sim Q_{BH,min}^2$, the extremality parameter simplifies to
\begin{equation}
\label{cextrhd2}
c^2=M_{BH,min}^2-Q_{BH,min}^2+2\alpha\, .
\end{equation}
For $c=0$ the minimal black hole is extremal, while for $c>0$ it is sub-extremal.

In general dimension $d$, the higher-derivative corrections are weighted by the UV cutoff $\Lambda$ according to
\begin{equation}
\frac{a_1}{\Lambda^d}(F_{\mu\nu}F^{\mu\nu})^2\, ,\qquad \frac{a_3}{\Lambda^2}F_{\mu\nu}F_{\rho\sigma}W^{\mu\nu\rho\sigma}\, ,
\end{equation}
where one expects $a_i = \mathcal{O}(1)$. Thus $\alpha_1 = \frac{M_{P}^d}{\Lambda^d} \, a_1$ and $\alpha_3 = \frac{M_P^2}{\Lambda^2} \, a_3$. Therefore, away from the species limit $\alpha_i = \mathcal{O}(1)$ and the extremality parameter is modified by a suitable linear combination $\alpha$. In the species limit the gravitational sector of higher-derivative corrections is expected to have $\Lambda \ll M_P$ and $\alpha_1 \gg \alpha_3 \gg 1$, unless there is a parametric suppression of the $a_i$, but in principle the electromagnetic sector considered above may have $\mathcal{O}(1)$ Wilson coefficients in Planck units. One way for this to occur is if the electromagnetic coupling $g$ is small, so that the canonical normalization in $\frac{1}{4} F_{\mu \nu} F^{\mu \nu}$ translates into factors of $g^2$ in the expressions for the Wilson coefficients\footnote{A more physical intuition behind such potential discrepancy is that the species contributing to electromagnetic higher-derivative corrections can be a small, charged, subset.}. At any rate, insofar as the linear combination $\alpha \geq 0$ black holes can satisfy the corrected extremality bound even when formed by super-extremal constituents, since the extremality parameter is corrected as \cite{Cheung:2018cwt}
\begin{equation}
\label{cextrhd_dim}
    c^2=M_{BH,min}^2-Q_{BH,min}^2+2\alpha \, M_{BH,min}^{2\frac{d-4}{d-3}} \, .
\end{equation}

\subsubsection{Non-extremal KK-like tower}
\label{sec:nonexKKtower}

As an example, let us consider a toy model with a tower of charged particles with equally spaces spectrum given as 
\begin{equation}
m_n=m_0 \, n \, ,\qquad n=1,\dots , N_{sp}\, .
\end{equation}
We assume that the electric charges of these particles are given by
\begin{equation}
q_n=m_0(n + \beta) \, ,\qquad n=1,\dots , N_{sp}.
\end{equation}
Enforcing the tower weak gravity conjecture, i.e.~that the entire tower be super-extremal, implies that $\beta\geq 0$.
Actually, to really have a conserved $U(1)$ gauge symmetry the tower must be extremal: it should not decay and $\beta=0$. 
In this case, we would also have that $T_{sp}=0$\footnote{Strictly speaking, this condition saturates the extremality bound at leading order in the large-charge expansion. Just like ordinary charged black holes, absent any (BPS or otherwise) protection, one expects that the extremality threshold be corrected by inverse powers of the charge. This is indeed what allows superextremal particle species to yield subextremal black holes in our setting.}.
However if the $U(1)$ gauge symmetry is mildly broken, the charges $q_n$ are not conserved anymore and the species can decay. 
In this case, $\beta\neq0$ and the species will possess a non-vanishing temperature. We will investigate the construction in string theory of this system in section \ref{sec:string_examples}.

We can see this explicitly by studying the total mass and charge of species. By summing over the tower up to $N_{sp}$, we get 
\begin{align}
M_{sp}&=\sum_{n=1}^{N_{sp}}m_n= {\cal S}_{sp}^{\frac{d-3}{d-2}}\left(1+{\cal S}_{sp}^{-\frac{1}{d-3}}+\dots\right)=\Lambda_{sp}^{3-d}+\Lambda_{sp}+\dots\, ,\nonumber\\
Q_{sp}&=\sum_{n=1}^{N_{sp}}q_n= {\cal S}_{sp}^{\frac{d-3}{d-2}}\left(1+(1+\beta){\cal S}_{sp}^{-\frac{1}{d-3}}+\dots\right)=\Lambda_{sp}^{3-d}+(1+\beta)\Lambda_{sp}+\dots\, ,
\end{align}
where we used $N_{sp}=\mathcal{S}_{sp}$ and dots represent terms that are subleading for large $N_{sp}$. 
For $d=4$, we can insert the above species mass and charge in the extremality bound including the contributions from the higher-derivative terms discussed above, namely relations \eqref{cextrhd1} and \eqref{cextrhd2}. Demanding $c\geq 0$ leads to
\begin{equation}
\alpha \gtrsim \beta.
\end{equation}
For $d>4$, in the species limit $\Lambda_{sp} \ll M_P$ the higher-derivative corrections to the extremality parameter scale as $\alpha \Lambda_{sp}^{2(4-d)}$ and dominate over the contributions coming from the tower, which scale as $\beta \Lambda_{sp}^{4-d}$. Then, at leading order in the species entropy the extremality parameter can be expressed as 
\begin{equation}
c\simeq\sqrt\alpha\Lambda_{sp}^{4-d} \simeq \sqrt\alpha({\cal S}_{sp})^{\frac{d-4}{d-2}}\, ,
\end{equation}
while the species temperature is
\begin{equation}
T_{sp}\simeq {\frac{c}{{\cal S}_{sp}}}={\frac{\sqrt\alpha}{ ({\cal S}_{sp})^{\frac{2}{d-2}}}}\simeq\sqrt\alpha (\Lambda_{sp})^{2}\, .
\end{equation}
One can check that all thermodynamic relations for charged black holes, respectively for charged species, are satisfied.
Note that, when compared to that of an uncharged tower in  \eqref{tspkk}, the temperature derived here in the charged case is suppressed by a factor of ${\cal S}_{sp}^{-\frac{1}{d-2}}$.

Finally, let us determine the quantum decay rate of the species particles. Assuming again that the charged species decay as quantum mechanical particles of mass $m_{sp}\simeq T_{sp}$ we get 
\begin{equation}
\Gamma_{sp,qm}\simeq T_{sp}^{d-1}\simeq \Lambda_{sp}^{2(d-1)}\, .
\end{equation}
In terms of entropy suppression, this corresponds to $n=2$ in \eqref{qmdecayRN}, i.e.~$\Gamma_{sp,qm}\simeq T_{sp}{\cal S}_{sp}^{-2}$.

\section{Comments on species cosmology and the Dark Dimension}

In this section, we collect some comments and observations on the role of species in cosmology in general and in the  Dark Dimension scenario \cite{Montero:2022prj} in particular. The main point we would like to make is that the thermodynamics of near-extremal species naturally suggests their possible use as dark matter candidates.

\subsection{Species decay in an expanding universe}

Towers of massive species may potentially play an interesting role in cosmology, both at early and late times. 
In particular, within the Dark Dimension scenario \cite{Montero:2022prj}, it was suggested that a tower of KK gravitons can play the role of dark matter particles \cite{Gonzalo:2022jac,Obied:2023clp}. 
To understand this proposal, is important to first discuss the production and decay of species in a universe expanding at a certain temperature $T_{univ}(t)$, as a function of time $t$. For concreteness, when considering near-extremal species, we will refer to the toy model in the preceding section, here applied to the KK modes of an appropriate internal space.

Recall that a tower of species is characterized by a thermodynamic temperature, $T_{sp}$, setting the maximal possible temperature the tower can acquire.
Therefore, it is reasonable to assume that a tower of species gets thermally produced at the time $t_i$ in the early universe when the temperature of the universe reaches that of the species, namely when
\begin{equation}
T_{univ}(t_i)\simeq T_{sp}\, .
\end{equation}
In the last sections we have discussed two different scenarios: neutral, Schwarzschild-like species, and near-extremal charged species. 
In terms of the species scale the corresponding temperatures in Planck units are
\begin{align}
\label{Tspneut}
{\rm Neutral ~Species:}\quad T_{univ}(t_i)&\simeq T_{sp}^{\rm neutral}\simeq\Lambda_{sp}\, , \\
\label{Tspnearext}
{\rm Near ~ extremal ~Species:}\quad T_{univ}(t_i)&\simeq T_{sp}^{\rm near-ext}\simeq \Lambda_{sp}^{2}\, .
\end{align}
We see that initial temperature of neutral species is as high as the UV cut-off scale, whereas the initial temperature of near-extremal species is parametrically smaller by a factor $\Lambda_{sp}$, and
\begin{equation}
 T_{sp}^{\rm near-ext}(t_i) \ll T_{sp}^{\rm neutral}(t_i).
 \end{equation}

At the moment of their production, the mass distribution of species, $m_{sp}(T)$, is thermal and peaked around  its maximum, $m_{sp}^{max}=T_{univ}(t_i)$.
Immediately after being produced, species start to  thermally decay with rate $\Gamma_{sp}\simeq T_{univ}(t_i)\simeq m_{sp}^{max}$. 
Actually, production of species and their subsequent thermal decay typically take place during the radiation epoch of the universe, when the energy density $\rho$ grows with temperature according to the $d=4$ Stefan's law, $\rho(t)\sim (T_{univ}(t))^4$, while the time-dependent scale factor of the universe, $a(t)\sim t^{1/2}$, grows with the inverse temperature as $a(t)\simeq (T_{univ}(t))^{-1}$. 
It follows that during the radiation epoch the expansion parameter $H(t)=\dot a(t)/a(t)$ is related to the temperature of the universe as
\begin{equation}
H(t)\simeq (T_{univ}(t))^2\, .
\end{equation}
Comparing then the expansion rate of the universe with the thermal decay rate of species at the time when they are produced, $\Gamma_{sp,th } $, we see that 
\begin{equation}
\Gamma_{sp,th}\simeq T_{univ}(t_i) > (T_{univ}(t_i))^2\simeq  H(t_i)\, .
\end{equation}
Therefore, species thermally decay with faster rate than the expansion of the universe. 
This process roughly lasts until half of the species have decayed. Then, as discussed previously, we expect species to acquire a quantum mechanical behavior and the fast thermal decay rate is replaced by the slower quantum mechanical one, which in $d=4$ behaves as 
$\Gamma_{sp,qm}\simeq T_{sp}^3$. Crucially, this is now slower than the expansion rate of the universe,
\begin{equation}
\Gamma_{sp,qm}\simeq T_{univ}^3 < T_{univ}^2\, \simeq H,
\end{equation}
which means that species decouple from expansion. 
From this moment on, the expansion of the universe takes place in a background with approximately constant UV cut-off $\Lambda_{sp}$.

\subsection{Species decay in the Dark Dimension scenario}

We apply now these general facts on the decay of species in an expanding universe to the Dark Dimension scenario.
The starting point is the Anti-de Sitter conjecture (ADC) \cite{Lust:2019zwm}, specialized to the case of positive vacuum energy density, which implies a UV/IR mixing between the UV cut-off $\Lambda_{sp}$ and the IR cosmological constant $\Lambda_{cc}$.
Concretely, the ADC states that in the limit of vanishing cosmological constant there must be a light tower of states, whose mass scale $m$ is parametrically related to $\Lambda_{cc}$ in the following way
\begin{equation}
\text{ADC}:\qquad m\simeq \Lambda_{cc}^\alpha\, M_P^{1-4\alpha}\, .
\end{equation}
The parameter $\alpha$ is bounded as 
\begin{equation}
{\frac{1}{ d}}\leq \alpha\leq {\frac{1}{2}}\, ,
\end{equation}
where the upper bound is model-independent and originates from the Higuchi bound \cite{Higuchi:1986py}, while the lower bound can be inferred by considering 1-loop Casimir potentials in $d$ dimensions \cite{Montero:2022prj, Anchordoqui:2023laz}.
Applying then constraints from experiments, one is 
led to the Dark Dimension Scenario \cite{Montero:2022prj} and its implications
\cite{Anchordoqui:2022ejw,Anchordoqui:2022txe,Blumenhagen:2022zzw,Gonzalo:2022jac,Anchordoqui:2022tgp,Anchordoqui:2022svl,Anchordoqui:2023oqm}
with $d=4$,  $\alpha=1/4$ and a tower of KK species coming from the compactification of one (large) extra dimension of size\footnote{An important correction factor $\lambda \lesssim 10^{-3}$ is understood in the subsequent relations.} 
\begin{equation}
R_{dd}\simeq \Lambda_{cc}^{-1/4}\simeq 1\mu{\rm m}\, .
\end{equation}
This length scale corresponds to a KK mass scale of order
\begin{equation}
m_{KK}\simeq \Lambda_{cc}^{1/4}\simeq 1~{\rm eV}\, 
\end{equation}
and it leads to a species scale 
\begin{equation}\
\Lambda_{sp}\simeq (\Lambda_{cc})^{\frac{1}{12}}\, .
\end{equation}
As for the temperatures of the KK species in terms of $\Lambda_{cc}$, from \eqref{Tspneut} and \eqref{Tspnearext} we get 
\begin{align}
{\rm Neutral ~Species:}\quad & T_{sp}^{\rm neutral} \simeq (\Lambda_{cc})^{\frac{1}{d(d-1)}}  \xrightarrow[d=4]{}  (\Lambda_{cc})^{\frac{1}{12}}\, , \\
\label{templambdacc}
{\rm Near ~extremal~Species:}\quad &T_{sp}^{\rm near-ext}\simeq (\Lambda_{cc})^{\frac{2}{d(d-1)}} \xrightarrow[d=4]{} (\Lambda_{cc})^{\frac16}\, .
\end{align}

We can now compare these temperatures with the mass scales relevant to realize the scenario in which the KK species play the role of dark matter candidates, as proposed in \cite{Gonzalo:2022jac}. 
There, it was argued that in order for the internal dimension to be sufficiently stabilized compared to the de Sitter mass scale, the initial temperature of the production of KK modes should be of order 
\begin{equation}
T_{univ}(t_i)\simeq (\Lambda_{cc})^{\frac{1}{2(d-1)}}\xrightarrow[d=4]{} \Lambda_{cc}^{1/6}\simeq 1~{\rm GeV}\, .
\end{equation}
This precisely agrees with the temperature \eqref{templambdacc} of the near-extremal species in $d=4$, which we argued to be the temperature and the mass scale at which species are produced during the radiation period of the expanding universe. Moreover, for phenomenological reasons it is important to point out that the charge we are considering in this scenario is associated with a new $U(1)_{\text{dd}}$ gauge symmetry possibly dynamically broken.

Hence, the thermodynamics of near-extremal species naturally provides the desired initial temperature at the time when the dark matter particles are produced. This result emphasizes the already acknowledged connection between dark matter as KK graviton, and black holes in the context of early time universe \cite{Anchordoqui:2022tgp}.

The decay of the dark matter KK particles after their production at $T_{univ}(t_i)$ can be constrained by the requirement of getting the correct dark matter abundance in the universe. 
Concretely, starting from a distribution that is initially peaked at around $T_{univ}(t_i)$, the bulk of the species masses has to shift down to about $1$-$100$ keV today. 
We can compare this value with the decay rate of the KK species. Right at the beginning, just after their production, these decay with a thermodynamic rate of order $\Gamma_{KK,th}\simeq T_{univ}(t_i)$, which is much faster than the required decay rate of dark matter particles. 
However, as argued before, after roughly half of the species life time, their decay is slowed down to the quantum mechanical rate $\Gamma_{KK,qm}\simeq T_{univ}(t_i)^3$. This is just what is needed to get the correct dark matter distribution \cite{Gonzalo:2022jac}. 
In conclusion, we argued that general facts of species thermodynamics naturally lead to the correct initial temperature and the correct decay rate required for near-extremal KK species to be viable dark matter candidates.

\section{Constructions in string theory}\label{sec:string_examples}

In this section, we explore constructions in string theory which can potentially provide a microscopic origin for the (near-)extremal species discussed above. Concretely, we will look at models in which supersymmetry is either broken \`a la Scherk-Schwarz or at the string scale $M_s^{-1} = 2\pi \sqrt{\alpha'}$. In both cases, the effects of supersymmetry breaking are controlled by a (small) parameter, which allows us to obtain light towers of super-extremal species, i.e.~with masses (slightly) smaller than charges.

\subsection{Kaluza-Klein towers in Scherk-Schwarz compactifications}\label{sec:KK_scherk-schwarz}

A natural starting point is the simplest Scherk-Schwarz compactification of type IIB string theory, where one reduces the theory on a circle with antiperiodic boundary conditions for fermions. From a field theory analysis one can notice that supersymmetry is spontaneously broken and that bosonic and fermionic Kaluza-Klein modes have different masses. This can be confirmed in string theory by computing the torus partition function, $\mathcal{T}$, whose integral over the fundamental domain yields the one-loop spacetime vacuum energy. The expression contains various sectors, as required by modular invariance, and (up to a prefactor) it takes the form \cite{Angelantonj:2002ct}
\begin{equation}\label{eq:scherk-schwarz_torus}
\begin{split}
\mathcal{T} & = |V_8 - S_8|^2 \frac{\sum_{m,n \in \mathbb{Z}} q^{\frac{\alpha'}{4}\left(\frac{m}{R} + \frac{n}{\tilde{R}}\right)^2} \, \bar{q}^{\frac{\alpha'}{4}\left(\frac{m}{R} - \frac{n}{\tilde{R}}\right)^2}}{ \eta \bar{\eta}} \\
& + |V_8 + S_8|^2 \frac{\sum_{m,n \in \mathbb{Z}}  (-1)^m q^{\frac{\alpha'}{4}\left(\frac{m}{R} + \frac{n}{\tilde{R}}\right)^2} \, \bar{q}^{\frac{\alpha'}{4}\left(\frac{m}{R} - \frac{n}{\tilde{R}}\right)^2}}{ \eta \bar{\eta}} \\
& + |O_8 - C_8|^2 \frac{\sum_{m,n \in \mathbb{Z}} q^{\frac{\alpha'}{4}\left(\frac{m}{R} + \frac{n+\frac{1}{2}}{\tilde{R}}\right)^2} \, \bar{q}^{\frac{\alpha'}{4}\left(\frac{m}{R} - \frac{n}{\tilde{R}}\right)^2}}{ \eta \bar{\eta}} \\
& + |O_8 + C_8|^2 \frac{\sum_{m,n \in \mathbb{Z}} (-1)^m q^{\frac{\alpha'}{4}\left(\frac{m}{R} + \frac{n+\frac{1}{2}}{\tilde{R}}\right)^2} \, \bar{q}^{\frac{\alpha'}{4}\left(\frac{m}{R} - \frac{n}{\tilde{R}}\right)^2}}{ \eta \bar{\eta}} \, .
\end{split}
\end{equation}
Here, the circle has radius $R$ and its T-dual radius is given by $\tilde{R} = \frac{\alpha'}{R}$. The affine characters $O_8, V_8, S_8, C_8$ are associated to the four conjugacy classes of the transverse isometry algebra $so(8)$, and correspond schematically to a scalar, a vector, and two spinors of opposite chirality, while the lattice sums comprise Kaluza-Klein and winding modes indexed by $m$ and $n$. We omitted the remaining factors of $\eta \bar{\eta}$ arising from the bosonic oscillators in the non-compact dimensions.

For sufficiently small radii, the $O_8 \bar{O}_8$ terms contain a level-matched tachyon, whose corresponding squared mass for $m=0$ and $n=0 \, , \, -1$ is given by
\begin{equation}
    m_{\text{tachyon}}^2 = - \, \frac{1}{2\alpha'} + \frac{R^2}{16{\alpha'}^2} \, .
\end{equation}
However, we are interested in the large $R$ limit, where this issue does not arise.\footnote{One subtle aspect of breaking supersymmetry is that $R$ is not a proper modulus, and it is generically subject to a force due to string-loop effects. Here we neglect this issue and focus on the nature of the light tower of wrapped D-branes.} In this limit only the first and the second line of \eqref{eq:scherk-schwarz_torus} keep a finite mass (for $n=0$) due to the twist in the winding modes. In particular, expanding the characters one sees that the bosonic combinations $V_8 \bar{V}_8$ and $S_8 \bar{S}_8$ contain the projector proportional to $(1 + (-1)^m)$, and thus only states with even $m$ are kept. Similarly, the combinations which are fermionic in spacetime keep only states with odd $m$. As a result, the Kaluza-Klein towers associated to massless states have masses $M = \frac{\abs{m}}{2R}$, with $\abs{m}=2k$ or $2k+1$ for bosons and fermions respectively so that one has
\begin{equation}
    m_{k,\text{bosons}} = \frac{k}{R} \, , \qquad m_{k,\text{fermions}} = \frac{k + \frac{1}{2}}{R} \, .
\end{equation}
As discussed before, compactifying on a circle introduces $U(1)$ charges coupled to the graviphoton, and one obtains extremal KK and winding towers, since the charge is the mass itself. In order to obtain a shift between mass and charge, we consider D1-branes wrapped on the circle, along the lines of \cite{Bonnefoy:2018tcp}. Because of supersymmetry breaking the tension is corrected, and for large radii $R \gg \sqrt{\alpha'}$ the correction can be estimated as $\delta T_{\text{D}1} \simeq - \frac{1}{2\pi R^2}$. The shift in the mass $m_{\text{D}1} = 2\pi R T_{\text{D}1}$ is thus additive and small in this limit, but the mass itself is large relative to the $9d$ Planck scale, since it is proportional to $R/g_s$.

In order to circumvent this issue and produce a tower lighter than the Planck mass, we further compactify five dimensions on a square torus of radius $L$. The $4d$ Planck scale is now $M_P \propto \sqrt{\frac{M_s^8 R L^5}{g_s^2}}$, and thus the ratio
\begin{equation}
    \frac{m_{\text{D}1}}{M_P} \propto (M_s R)^{\frac{1}{2}} (M_s L)^{-\frac{5}{2}} \left( 1 + \frac{g_s}{2\pi} \, \frac{\delta T_{\text{D}1}}{M_s^2}\right)
\end{equation}
is bounded as $g_s \to 0$. Taking for instance $L=R \gg \sqrt{\alpha'}$ the correction $\delta T_{\text{D}1}$ simplifies as in \cite{Bonnefoy:2018tcp}, and thus
\begin{equation}
    \frac{m_{\text{D}1}}{M_P} \sim \frac{1}{M_s^2 R^2} \left(1 - \frac{g_s}{4\pi^2 M_s^2 R^2}\right) \, .
\end{equation}
In the above expression, one can also read off the winding charge $q_\text{D1} = \frac{1}{M_s^2 R^2}$, with respect to which these states are super-extremal. Otherwise, one can also keep $R$ string-sized (but still above the tachyon threshold). In this case, the shift $\delta T_{\text{D}1}$ is $\mathcal{O}(M_s^2)$ and can be numerically extracted, for instance, from the tree-channel annulus amplitude $\widetilde{\mathcal{A}}_{11}$ evaluated at zero separation between the branes.
In either case, there are also light extremal KK modes, and for $L = R \gg \sqrt{\alpha'}$ they are lighter than the wrapped D1-branes by a factor of $\frac{g_s}{M_s^2R^2}$. It seems plausible that the correction to their masses due to supersymmetry breaking, if any, be multiplicative rather than additive, since in the large-radius limit the effect ought to be captured by a field theory analysis of gravitational interactions.

\subsection{Non-supersymmetric brane configurations}\label{sec:non-susy_branes}

In light of the emergent string conjecture  \cite{Lee:2019tst, Lee:2019wij}, which holds that the lightest towers of species in string theory arise only as Kaluza-Klein towers or excitations of weakly coupled strings, another interesting limit to consider is weak string coupling $g_s \to 0$ in (orientifold or heterotic) models where supersymmetry is broken at the string scale. The simplest examples, which have no tachyons in ten dimensions, are the Sugimoto \cite{Sugimoto:1999tx} and Sagnotti \cite{Sagnotti:1995ga, Sagnotti:1996qj} orientifold models and the unique such heterotic model \cite{Alvarez-Gaume:1986ghj, Dixon:1986iz}. These are characterized, among other features, by an exponential potential,
\begin{equation}\label{eq:dilaton_tadpole}
    \delta S_\text{SUSY-breaking} = - \frac{M_s^8}{2} \int d^{10}x \, \sqrt{-g} \, T \, e^{\chi \phi},
\end{equation}
arising due to supersymmetry-breaking effects. Here,  $\chi = -1 \, , \, 0$ for the orientifold and heterotic models (in string frame), encoding the leading contribution due to annulus and torus amplitudes respectively.\footnote{For recent reviews of this and related aspects of non-supersymmetric models, see \cite{Mourad:2017rrl,Basile:2021vxh,Sagnotti:2021mxb}.} 
The fixed and calculable constant $\alpha' T$ is of order one, as befits the breaking of supersymmetry at the string scale, but its precise value is irrelevant for our purposes.

In such settings, weakly coupled (and possibly long-lived) vacua can be constructed \cite{Mourad:2016xbk} with suitable brane configurations, and supersymmetry breaking can renormalize and decrease the effective tension $T$ of branes relative to their charge $q$ \cite{Basile:2018irz, Antonelli:2019nar, Basile:2021mkd, Basile:2021vxh}. This renormalization is multiplicative, of the form $T_\text{eff} = q / (1 + \delta)$, rather than additive as in the preceding settings. Specifically, the examples where the supersymmetry-breaking effects in the backreacted geometries are known correspond to D1-branes and D3-branes in the orientifold models, and NS5-branes in the heterotic model. For D1-branes and NS5-branes the multiplicative shift is $\delta_\text{D1} = \sqrt{\frac{3}{2}} - 1$ and $\delta_\text{NS5} = \sqrt{\frac{5}{3}} - 1$ respectively, while for D3-branes one finds $\delta_\text{D3} \propto g_s^2 N \, \log u \ll 1$, which depends on the string coupling, the number $N$ of D3-branes sourcing the geometry, and the holographic coordinate $u$ describing the transverse distance to the sourcing stack (at least within the regime of validity of the EFT computation). As in the preceding section, this means that compactifying to four dimensions on a square torus of radius $R \gg \sqrt{\alpha'}$ and wrapping D1-branes on a 1-cycle yields particles whose masses are multiplicatively corrected according to
\begin{equation}
    \frac{m_{\text{D}1}}{M_P} \sim \sqrt{\frac{2}{3}} \, \frac{1}{M_s^2R^2} = \sqrt{\frac{2}{3}} \, q_{\text{D}1} \, .
\end{equation}
Without compactifying, examples of such constructions so far describe branes that are heavy as $g_s \to 0$, which is not suitable for our purposes.

In order to see which objects become light in the limit of weak string coupling, let us consider a charged probe $p$-brane with a generic string frame action
\begin{equation}
    S_\text{probe} = - T \int d^{p+1}\xi \, \sqrt{-g} \, e^{- \sigma \, \phi} + q \int C_{p+1},
\end{equation}
where the pullback to the worldvolume of the bulk fields is implicit. The coupling to the dilaton $\phi$ is characterized by $\sigma$, which is 0 for F-strings, 1 for D-branes and 2 for NS-branes. Passing to Einstein frame in $d$ spacetime dimensions, the effective tension reads
\begin{equation}
    T_\text{eff} = T \, e^{\left( 2\frac{p+1}{d-2} - \sigma \right) \phi} \, ,
\end{equation}
and thus, as $g_s = e^\phi \to 0$, the brane is light whenever $p+1 > \frac{d-2}{2} \, \sigma$. Barring exotic branes along the lines of \cite{Bergshoeff:2005ac, Bergshoeff:2006gs, Bergshoeff:2011zk, Bergshoeff:2012jb}, the only possibilities in $d=10$ are F-strings and D$p$-branes with $p>3$. In order to be able to obtain charged black holes in four or more extended dimensions, the only possibilities left in these models are thus F-strings in the heterotic model and D5-branes in either orientifold model. These are actually not included in the cases listed above, but fortunately the effects of supersymmetry breaking are subleading in the near-horizon limit for these objects \cite{Bogna:2021the}. Thus, one can still study the effective tension of probe branes, organizing the expansion around an extremal solution with radius $R$ (which determines the ADM tension and charge) in powers of 
\begin{equation}
    \rho \equiv T R^2 g_s^\gamma \ll 1 \, ,
\end{equation}
where $\gamma = 3/2$ ($5/2$) for the orientifold (heterotic) case is the Einstein-frame counterpart of the exponent in \eqref{eq:dilaton_tadpole}.

Computing the leading corrections to the supersymmetric solution and the corresponding potential felt by a probe, one obtains the following results from \cite{Bogna:2021the}.

For D5-branes, it turns out that the probe potential vanishes even to linear order in the supersymmetry-breaking parameter $\rho$. Therefore, the simplest setting left to consider is given by F-strings in the heterotic model, where the leading corrections to their backreaction takes a simple form \cite{Bogna:2021the}. The near-horizon backreacted geometry due to a stack of straight, coincident and extremal F-strings in Einstein frame is corrected at leading order in $\rho$ according to
\begin{align}
    ds^2 & = \left[\left(\frac{r}{R}\right)^{\frac{9}{2}} + \frac{107}{8960} \, \rho \left( \frac{r}{R} \right)^{\frac{25}{2}}\right] dx_{1,1}^2 + \left[\left(\frac{r}{R}\right)^{-\frac{3}{2}} - \frac{45}{8960} \, \rho \left( \frac{r}{R} \right)^{\frac{13}{2}}\right] \left(dr^2 + r^2 \, d\Omega_7^2 \right) \, , \\
    e^\phi & = g_s \left[\left( \frac{r}{R} \right)^3 + \frac{356}{8960} \, \rho \left( \frac{r}{R} \right)^{11}\right] \, , \qquad B_2 = \sqrt{g_s} \left[ \left(\frac{r}{R}\right)^6 + \frac{9}{224} \, \rho \left(\frac{r}{R}\right)^{14} \right] \, dx \wedge dt \, .
\end{align}
Evaluating the Nambu-Goto action including the coupling to the B-field for an F-string probing this background,
\begin{equation}
    S_\text{F1} = - \frac{1}{2\pi \alpha'} \int d^2x \, \sqrt{-g} \, e^{\frac{\phi}{2}} + \frac{1}{2\pi \alpha'} \int B_2 \, ,
\end{equation}
one obtains the leading-order potential
\begin{equation}
    V_{\text{F1}} = \frac{\sqrt{g_s}}{2\pi \alpha'} \, \rho \left(\frac{107}{8960} + \frac{1}{2} \, \frac{356}{8960} - \, \frac{9}{224} \right) \left(\frac{r}{R}\right)^{14} = - \, \frac{\sqrt{g_s}}{2\pi \alpha'} \, \frac{15\rho}{1792} \left(\frac{r}{R}\right)^{14} \, ,
\end{equation}
where we highlighted the contributions from the metric warping, the dilaton and the B-field in this order. This potential is repulsive, reflecting the fact that the effective tension-to-charge ratio of the F-string due to the supersymmetry-breaking effects is renormalized to
\begin{equation}
    \left( \frac{T}{q} \right)_\text{eff} = \frac{1 + \frac{107}{8960} + \frac{1}{2} \, \frac{356}{8960}}{1 + \frac{9}{224}} = \frac{1849}{1864} \approx 0.99 \, .
\end{equation}
These strings can be wrapped on small cycles to give rise to slightly subextremal particles with a multiplicative shift in the extremality ratio. In order to take into account the presence of the dynamical tadpole, the simplest setting to consider is given by Dudas-Mourad vacua.

\subsubsection{F-strings in Dudas-Mourad vacua}

According to the above results, one may be tempted to build quasi-extremal (species) black holes using fundamental strings in the Sagnotti model. In order to produce black holes rather than black strings, naively one ought to hide the spatial extent of the strings in some internal dimensions. The presence of the dilaton tadpole makes it difficult to stabilize the internal dimensions in a controlled fashion leaving an asymptotically flat extended spacetime. However, there is a workaround, with the proviso of giving up parametric control globally. The effective field equations with the inclusion of the tadpole potential \eqref{eq:dilaton_tadpole} afford solutions with flat spacetime fibered over an interval, along which the dilaton also runs. The simplest solution of this type was found by Dudas and Mourad \cite{Dudas:2000ff}, and its dimensional reduction over this interval was recently revisited in \cite{Basile:2022ypo}. Similar solutions in lower dimensions, reduced over tori, have been thoroughly analyzed in \cite{Mourad:2021roa, Mourad:2022loy, Mourad:2023loc}. The simplest setting to obtain a black hole in the nine-dimensional Dudas-Mourad reduction is simply to place the string along the internal interval. However, the endpoints of the interval feature strongly coupled and/or curved regions, and it is unclear to which extent these affect the reliability of the solution. For instance, it has been argued that, while perturbatively stable, the Dudas-Mourad geometry may be affected by strong-coupling instabilities due to the endpoints of the interval \cite{Raucci:2023xgx}. 
Another setup, which avoids placing the black string in the problematic region, involves reducing the geometry (say on a torus) to four dimensions, wrapping the black string on an internal cycle. This results in a black hole in a geometry fibered over an interval of length proportional to $g_s^{- \gamma/2}$. Reducing to four external dimensions, this scenario would look somewhat similar to black holes in the dark dimension, where the Dudas-Mourad interval plays the role of the mesoscopic extra dimension. However, in this setting spacetime would remain flat, and additional ingredients are required to produce a quasi-de Sitter scenario.

\section{Conclusion}

In this work, we extended the notion of species thermodynamics \cite{Cribiori:2023ffn} along various directions.
First, by exploiting the N-portrait description of minimal black holes, we made the correspondence between towers of species and minimal black holes more concrete.
We started from the case of neutral (uncharged) towers, such as KK towers originating from compactification on a certain compact space. 
Here, we could show explicitly that counting the microstates of the KK bound state precisely matches with the thermodynamic entropy of the corresponding minimal black hole.
It follows that the thermodynamic black hole relations between mass, temperature and entropy are fulfilled for a tower of KK modes.\footnote{The question what kind of species tower satisfies thermodynamic black hole relations and an associated bottom-up black hole derivation of the emergent string conjecture was recently investigated in more detail in \cite{Basile:2023blg}.} Next we extended the N-portrait picture for charged species and charged minimal black holes. 
With respect to the uncharged case, the thermodynamic relations are modified, with the result the temperature of charged, near-extremal species is suppressed compared to their neutral counter parts. 
This has interesting consequence on the behavior of near-extremal towers of species in the early universe. In particular, in an expanding universe charged species are more stable and do not decay as fast as neutral species. In particular, we find a possible way to explain the initial temperature of KK gravitons \cite{Gonzalo:2022jac}, as near-extremal black holes.
In the last section we considered how near-extremal species can possibly be obtained in string and brane constructions with (spontaneously) broken supersymmetry. Concretely, we looked at KK species in Scherk-Schwarz compactifications and at towers of species arising from various (wrapped) strings and branes configurations in the non-supersymmetric and tachyonic-free orientifold and heterotic models. 
These results are summarized in Table \ref{tablesusy}.


\begin{table}[ht]
\begin{tabular}{lccc}
 & Mass in $4d$ Planck units & Mass-to-charge shift \\
 D1 (Scherk-Schwarz) & light & additive, $\ll 1$ \\
 F1 (heterotic) & light & multiplicative, $\approx 10^{-2}$\\
 D1 (orientifold) & light & multiplicative, $\mathcal{O}(1)$ \\
 NS5 (heterotic) & heavy & multiplicative, $\mathcal{O}(1)$ \\
\end{tabular}
\caption{Comparison of various candidates for near-extremal species black holes in non-supersymmetric string constructions. Our ideal target are particle-like light species with a parametrically small additive shift in mass relative to the charge. The most promising candidate are wrapped D1-branes on large circles in Scherk-Schwarz settings. Configurations of this type have soft supersymmetry breaking, but also dynamical tadpoles.}
\label{tablesusy}
\end{table}

Our findings can be extended along various directions. It would be important to improve further the thermodynamic picture of species and see to what extent it can give a rationale behind swampland conjectures, at the boundary but possibly even in the interior of the moduli space, which remains far less explored. It would be also interesting to explore the consequence of species thermodynamics in cosmology and possibly determine general predictions of phenomenological interest from theoretical considerations. More quantitative, it would be of clear importance to provide a concrete microscopic embedding of the scenarios here investigated, also in relation to the Dark Dimension. We hope to come back to these questions in the future.

\section*{Acknowledgements}

We thank  Cumrun Vafa and Timo Weigand  for insightful discussions. 
The work of D.L. is supported by the Origins Excellence Cluster and by the German-Israel-Project (DIP) on Holography and the Swampland.

\appendix

\section{Species entropy from graviton propagator}

In this appendix, we provide a derivation of the species entropy which does not proceed via black holes and it generalizes the known quantum field theory argument to the case of a CFT which is not necessarily weakly interacting.

One of the standard arguments supporting the existence of a species scale $\Lambda_{sp}<M_P$ makes use of perturbative quantum field theory. As reviewed for example in \cite{Castellano:2022bvr}, by looking at the corrections to the graviton propagator in a theory with $N_{sp}$ scalars, one finds that the one-loop correction is of the same order as the tree level term when $\Lambda_{sp} = M_P/N_{sp}^{\frac{1}{d-2}}$.\footnote{If species are strictly massless, one finds also a multiplicative logarithmic correction whose role is unclear. We believe that it is an artefact of the massless case and it shall not appear when species are massive \cite{Blumenhagen:2023yws}.} This argument typically relies on the species being weakly coupled to gravity but also amongst themselves. We argue now that the latter assumption can be removed. At the same time, within a CFT which is not necessarily weakly coupled, we will connect $N_{sp}$ to the species entropy introduced in \cite{Cribiori:2023ffn}.

Let us denote with $h_{\mu\nu}$ the linearized graviton and with $T_{\mu\nu}$ the species stress tensor. Assuming that the coupling $h_{\mu \nu} T^{\mu \nu}$ is small (without imposing that species are weakly interacting among themselves), we can compute the leading correction to the graviton propagator. Second-order perturbation theory for gravitons reveals that such correction is controlled by the Fourier transform $\mathcal{F}$ of the (Euclidean) matter correlator \cite{Anselmi:1997rd}
\begin{equation}
\langle T_{\mu \nu}(x) T_{\rho \sigma}(0)\rangle = \pi_{\mu \nu \rho \sigma} \, \frac{c(\abs{x})}{\abs{x}^4} + \pi_{\mu \nu} \, \pi_{\rho \sigma} \, \frac{f(\abs{x})}{\abs{x}^4} \, .
\end{equation}
Here $c$, $f$ are dimensionless functions of their argument, while $\pi_{\mu \nu} \equiv \partial_\mu \partial_\nu - \delta_{\mu \nu} \Box$ and $\pi_{\mu\nu\rho\sigma} = 2\pi_{\mu\nu}\pi_{\rho\sigma} - 3(\pi_{\mu\rho}\pi_{\nu\sigma}+\pi_{\mu\sigma}\pi_{\nu\rho})$ is the polarization tensor. 

When contracting with external gravitons one can therefore focus on the first term, and one obtains the relative correction
\begin{equation}
\frac{p^2}{M_P^2} \, \mathcal{F}\left(\frac{c(\abs{x})}{\abs{x}^4}\right)(p) \, ,
\end{equation}
which, up to numerical factors, gives
\begin{equation}\label{eq:integral_correction}
\frac{p^2}{M_P^2} \, \int_{\epsilon}^\infty \frac{dr}{r} \, c(r) \, \frac{J_1(\abs{p}r)}{\abs{p}r} =\frac{p^2}{M_P^2} \,  \int_{\epsilon\abs{p}}^\infty \frac{ds}{s^2} \, J_1(s) \, c\left(\frac{s}{\abs{p}}\right),
\end{equation}
where $J_1$ is the Bessel function of the first kind. 
Notice that we introduced an UV cutoff $\epsilon$ which in principle may be identified with the species scale itself. More generally, the species scale gives an upper bound on the ultraviolet cutoff of gravitational effective theories. 
For a CFT $c$ is a constant and, in the weakly coupled case, it reduces to $N_{sp}$ at leading order\footnote{This case was actually already considered in \cite{Dvali:2007wp}.}. Expanding the integral \eqref{eq:integral_correction} for small $\epsilon$, we get schematically
\begin{equation}
\frac{p^2}{M_P^2} c\int_{\epsilon\abs{p}}^\infty \frac{ds}{s^2} \, J_1(s) \simeq \frac{p^2}{M_P^2} c\left( const - 2\log(\epsilon|p|) + \mathcal{O}((\epsilon|p|)^2)\right)
\end{equation}
and by choosing $\epsilon \simeq 1/M_P$ we reproduce the known term $-c \, \frac{p^2}{M_P^2} \,  \log \frac{p^2}{M_P^2}$ \cite{Calmet:2017omb}. By asking that this be order one, namely comparable to the tree level term, one recovers the species scale (or more precisely an upper bound), up to a multiplicative logarithmic correction whose role is unclear. Since in no point we assumed the CFT to be weakly coupled, the above calculation is valid also in the strongly coupled case where $c$ is not necessarily $N_{sp}$. Rather, the function in \eqref{eq:integral_correction} can be interpreted as an effective number of (interacting) species $N_\text{eff}$.

To make the connection with the species entropy \cite{Cribiori:2023ffn}, let us recall that in a CFT at a temperature $T$ and in a volume $V$ the entropy is $\mathcal{S} \simeq c \, V \, T^3$. Then, at the minimal volume $V \simeq \Lambda_{sp}^{-3}$ and at the species temperature $T_{sp} \simeq \Lambda_{sp}$ one finds $\mathcal{S}_{sp} \simeq c$. This result thus generalizes the expression for $\Lambda_{sp}$ in terms of the number of species when their dynamics is strongly coupled. However, the integral expression in \eqref{eq:integral_correction} is more general and can take into account massive towers of degrees of freedom as well, and may provide a more concrete connection with the general proposal for the species scale away from weakly coupled limits. Nevertheless, \eqref{eq:integral_correction} gives different subleading corrections to the species scale for weakly coupled species, which can be compared to the usual definition.

For instance, when considering critical type IIA string theory one has that the tree-level contribution to $c$ at weak string coupling is in fact the number of species $N_{sp} \sim g_s^{-2}$ \cite{Dvali:2009ks}. Then, one can expect the presence of a subleading term at string one-loop. Including this correction, $c \sim g_s^{-2} + \alpha$ for some constant $\alpha$, and thus $\Lambda_{sp}^{-8} \propto  g_s^{-2}+\alpha$ in ten-dimensional Planck units. By comparison with the expression proposed in \cite{vandeHeisteeg:2023dlw,Castellano:2023aum}, $\Lambda_{sp}^{-6} \propto g_s^{-3/2} + \beta \, g_s^{1/2}$, one identifies $\alpha = \frac{4}{3} \, \beta$, with $\beta=\pi^2 \zeta(3)/3$. However, the above definition of species scale could in general differ from the one encoded in the typical higher-derivative corrections to the Einstein-Hilbert term, where with typical we mean those corrections suppressed with $\Lambda_{sp}$ in such a way to have order-one Wilson coefficients. In general, some operators may have fine-tuned coefficients.

\bibliography{references}
\bibliographystyle{JHEP}

\end{document}